\documentstyle[twoside,fleqn,espcrc2]{article}

\newcommand{\AmS}{{\protect\the\textfont2
  A\kern-.1667em\lower.5ex\hbox{M}\kern-.125emS}}

\title{Isospin Symmetry Breaking and the $\rho$--$\omega$--System}

\author{H. Fritzsch and A.S. M\"uller\address{Sektion Physik,
        Ludwig--Maximilians--Universit\"at M\"unchen, \\
        Theresienstrasse 37, D--80333 M\"unchen}}%
\begin{document}
\begin{abstract}
Simple quark models for the low lying vector mesons suggest a
mixing between the $u$-- and $d$--flavors and a violation of
the isospin symmetry for the $\rho-\omega$ system much stronger
than observed. It is shown that the chiral dynamics, especially
the QCD anomaly, is responsible for a restoration of the isospin
symmetry in the $\rho-\omega$ system.
\end{abstract}
\maketitle

We should like to report on an interesting phenomenon, thus far
unrecognized, which we found in the vector meson sector and which
is directly related to the violation of isospin symmetry and its
description whithin the QCD framework. It points towards a
dynamical restoration of isospin symmetry in the low energy
sector of QCD.\\
Although there are no doubts that all observed strong interaction
phenomena can be described within the theory of QCD, a quantitative
description of the strong interaction phenomena in the low energy
sector (e.g in the energy range $0...2$ GeV) is still lacking,
although some features of the low energy phenomena have been
partially understood by the lattice gauge theory approach.
Nevertheless a number of features of the strong interaction
phenomena at low energies can be related to basic symmetries like
isospin or $SU(3)$, to symmetry breaking effects and to basic
properties derived in simple phenomenological models.\\
The low energy sector of the physics of the strong interactions
is dominated by the low--lying pseudoscalar mesons
($\pi, K,\eta,\eta'$) and the low--lying vector mesons
($\rho,\omega, K^{*},\phi$). It is well--known that the structures of
the quark wave functions of the pseudoscalar mesons ($0^{-+}$) and of
the vector mesons ($1^{--}$) differ substantially.\\
In the vector meson channel there is a strong mixing between           
the eights component of the $SU(3)$ octet (wave function:
$(\bar u u +\bar d d -2\bar s s)/(\sqrt 6)$) and of the $SU(3)$
singlet (wave function: $(\bar u u +\bar d d +\bar s s)/(\sqrt 3)$).
The mixing strength is such that the mass eigenstates are nearly
the state $(\bar u u +\bar d d)/(\sqrt 2)$), the $\omega$--meson,
and the state $\bar s s$, the $\phi$--meson. While this feature looks
peculiar, when viewed upon from the platform of the underlying $SU(3)$
symmetry, it finds a simple interpretation, if one takes into account
the Zweig rule \cite{okub63}, which states that the
mixing must take place
in such a way that quark lines are neither destroyed nor created.\\
On the other hand the pseudoscalar mesons follow the pattern
prescribed by the $SU(3)$ symmetry in the absence of singlet--octet
mixing. The neutral mass eigenstates $\eta$ and $\eta'$ are nearly
an $SU(3)$--octet or $SU(3)$--singlet:
\begin{eqnarray}
\quad\eta &\approx& \frac{1}{\sqrt 6}\,(\bar u u +\bar d d -2\bar s s)
\quad\mbox{or}\\
\quad \eta' &\approx&\frac{1}{\sqrt
3}\,(\bar u u +\bar d d + \bar s s)\,. \nonumber
\end{eqnarray}
This indicates a large violation of the Zweig rule in the $0^{-+}$
channel
\cite{fritzsch75} \cite{venez89}. Large transitions between the various
($\bar q q$)--configurations
must take place. In QCD the strong mixing effects are related to the
spontaneous breaking of the chiral $U(1)$ symmetry normally attributed
to
instantons. Effectively the mass term for the pseudoscalar mesons can
written as
follows, neglecting the effects of symmetry breaking in the gluonic
mixing term \cite{fritzsch73} \cite{hooft76} \cite{shifm86}:
\begin{equation}
\quad M^2_{\bar q q} = \left( \begin{array}{ccc}
M^2_u & 0 & 0\\ 0 & M^2_d & 0\\ 0 & 0 & M^2_s \end{array} \right)\,
\end{equation}
\begin{displaymath}
\quad\quad\quad\quad\quad+\,\,
\lambda \left( \begin{array}{ccc} 1&1 &1\\ 1&1 &1\\ 1&1 &1\\            
 \end{array} \right)\,,
\end{displaymath}
where $M^{2}_{u},M^{2}_{d}$ and $M^{2}_{s}$ are the $M^{2}$--values
of the masses of quark composition $\bar u u, \bar d d$ and $\bar s s$
respectively.\\
It is well--known that the mass and mixing pattern of the
$0^{-+}$--mesons is described by such an ansatz \cite{fritzsch75}.
The parameter $\lambda$, which describes the mixing strength due to the
gluonic forces, is essentially given by the $\eta'$--mass:
$\lambda \cong 0.24$ GeV$^{2}$. Since $\lambda$ is large
compared to the strength of $SU(3)$ violation given by the $s$--quark
mass, large mixing phenomena are present in the $0^{-+}$ channel, as
seen in the corresponding wave functions.\\
The situation is different in the vector meson $1^{--}$ channel.
Here the gluonic mixing term is substantially smaller than the strength
of $SU(3)$ violation such that the Zweig rule is valid to a good
approximation. If one describe the mass matrix for the vector mesons
in a similar way as for the pseudoscalar, we have
\begin{equation}
M_{\bar q q} = \left( \begin{array}{ccc} M(\bar u u) & 0 & 0\\ 0 &
M(\bar d d) & 0\\ 0 & 0 & M(\bar s s) \end{array}\right)\,
\end{equation}
\begin{displaymath}
\quad\quad\quad\quad\quad\quad\quad\quad\,
+\,\,\tilde \lambda \left( \begin{array}{ccc} 1&1 &1\\
1&1 &1\\ 1&1 &1\\ \end{array} \right)\,,
\end{displaymath}
here $M(\bar q q)$ denotes the mass of a vector meson with quark
composition $\bar q q$ in the absence of the
mixing term. The magnitude of the mixing term $\tilde \lambda$ can be
obtained in a number of different ways, e.g by considering the
$\rho_{0}$--$\omega$ mass difference.
Neglecting the isospin violation caused by the $m_{d}$--$m_{u}$
mass splitting, the gluonic mixing term is responsible for the
$\rho_{0}$--$\omega$ mass shift:
\begin{equation}
\quad M_{\omega}-M_{\rho}= 2 \tilde \lambda\quad ,
\end{equation}
\begin{displaymath}
\quad\tilde \lambda \cong 6.0 \pm 0.5             
\,\,\mbox{MeV}\,.
\end{displaymath}
The decay $\phi\rightarrow 3\pi$ proceeds via the
$(\bar u u+\bar d d)$--admixture in the $\phi$ wave function.
Using the observed branching ratio
\begin{equation}
\quad\,\,\frac{\Gamma (\phi\rightarrow 3\pi)}{\Gamma (\omega\rightarrow
3\pi)}\,\,\simeq 0.09 \,,
\end{equation}
on finds a gluonic mixing term of the same order of magnitude.\\
In QCD the isospin symmetry is violated by the mass splitting
between the $u$-- and $d$--quark. Typical estimates give:
\begin{equation}
\quad \frac{m_{d}-m_{u}}{\frac{1}{2}(m_{d}+m_{u})}\cong 0.58\,.
\end{equation}
The observed smallness of isospin breaking effects is usually attributed
to
the fact that the mass difference $m_{d}$ -- $m_{u}$ is small compared
to the QCD scale $\Lambda_{QCD}$. However in the case of the vector
mesons
the QCD interaction enters in two different ways:\\\\
a) \, In the chiral limit of vanishing quark masses the masses of the
vector mesons are solely due to the QCD interaction, i.e.
$M= \mbox{const}\cdot\Lambda_{QCD}$.\\\\
b)\, The QCD mixing term will lead to a mixing among the various flavour

components such that the $SU(3)$ singlet (quark composition $(\bar u u+
\bar d d +\bar s s)/\sqrt 3$) is lifted upwards compared to the two
other neutral components given by the wave functions $(\bar u u - \bar d
d)/\sqrt 2$ and $(\bar u u + \bar d d-2 \bar s s)/\sqrt 6$. The
corresponding mass shift is given by $3\tilde \lambda$.\\\\
We approach the real world by first introducing the mass of the strange
quark. As soon as $m_{s}$ becomes larger than $3\tilde \lambda$,
substantial
singlet--octet mixing sets in, and the mass of one vector meson
increases until it reaches the observed value of the $\phi$--mass.
At the same time the Zweig rule, which is strongly violated in the        
chiral
$SU(3)_{L}\times SU(3)_{R}$ limit becomes more and more valid.\\
The validity of the Zweig rule is determined by the ratio
$m_{s}/\tilde\lambda$. If this ratio vanishes, the Zweig rule is
violated strongly. In reality, taking $m_{s}$ (1GeV) $\approx$ 150 MeV,
the ratio $m_{s}/\tilde\lambda$ is about $25$ implying that the
Zweig rule is nearly exact.\\
In a second step we introduce the light quark masses $m_{u}$ and
$m_{d}$.
We concentrate on the non--strange vector mesons. If the gluonic
mixing interaction were turned off, the mass eigenstates would be
$v_{u}=|\bar u u\rangle$ and $v_{d}=|\bar d d\rangle$. The masses of
these mesons are given by:
\begin{equation}
\quad M(v_{u}) = \langle v_{u}|\, H^{0}+ m_{u}\,\bar u u\,|
\,v_{u}\rangle\,,
\end{equation}
\begin{displaymath}
\quad M(v_{d}) = \langle v_{d}|\, H^{0}+ m_{d}\,\bar d d\,|
\,v_{d}\rangle\,.
\end{displaymath}
Here $H^{0}$ is the QCD--Hamiltonian in the chiral limit
$m_{u}=m_{d}=0$. We have assumed, as expected in simple valence quark
models that the matrix elements $\langle v_{u}|\,\bar d d\,|\,
v_{u}\rangle$
and $\langle v_{d}|\, \bar u u\,|\,v_{d}\rangle$ are very small and can
be neglected. Thus the masses can be written as
\begin{equation}
\quad M(v_{u}) = M_{0}+ 2m_{u}\cdot c\,,
\end{equation}
\begin{displaymath}
\quad M(v_{d})= M_{0}+ 2m_{d}\cdot c\,.
\end{displaymath}
(c: constant, given by the expectation value of $\bar q q$).
The introduction of the light quark masses induces positive mass
shifts for both $v_{u}$ and $v_{d}$. These mass shifts can be estimated
by considering the corresponding mass shifts of the charged
$K^{*}$--mesons:
\begin{eqnarray}
\Delta \tilde M & = & \tilde M(K^{*0})- \tilde M(K^{*+})\\       
& = & (m_{d}-m_{u})\cdot c = 4.44\,\,\mbox{MeV}\, , \nonumber
\end{eqnarray}
where $\tilde M$ is the mass of the vector meson in the absence of
electromagnetism. Taking the electromagnetic mass shift into account
\cite{scadr84}.
\begin{eqnarray} \Delta M^{2}(K^{*})_{elm}&\cong&
\frac{2}{3}\Delta M^{2}(\rho)\\
\Delta M^{2}(\rho)&=& \Delta M^{2}(K^{*})- 3\Delta M^{2}(K) \nonumber\\
& & +\,\frac{9}{2} \Delta M^{2}(\pi)\, , \nonumber
\end{eqnarray}
we find $M(K^{*})_{elm}\cong -3,59 $ MeV and
$\Delta \tilde M = (m_{d}-m_{u})\cdot c = 0.85$ MeV. Thus we
obtain for the mass shift of the neutral vector mesons:
\begin{eqnarray}
M(v_{d})-M(v_{u}) & \cong & 2\,(m_{d}-m_{u})\cdot c \\
& \cong & 1.7 \,\,\mbox{MeV}. \nonumber
\end{eqnarray}
We like to emphasise that our way of relating the mass differences
between the $(\bar u u)$ and $(\bar d d)$ vector mesons to the
mass differences between the $(\bar u s)$ and $(\bar d s)$ vector mesons

is more than using isospin symmetry, since the first two mesons are
members of an isotriplet, while the second two mesons form an
isodoublet.
Using simple $SU(6)$ type quark models or using $SU(3)$ symmetry with
the additional input that the $\bar q q$--operator has a pure
F--coupling,
in accordance with observation in the case of the baryons, the two mass
terms are indeed related, as we stated, i.e.
$M(\bar d d)-M (\bar u u)= 2(M(\bar s d)- M(\bar s u))$.\\
It is remarkable that this mass shift is of similar order of magnitude
as
the mass shift  between the isosinglet and isotriplet state in the
chiral
limit, where
isospin symmetry is valid. This implies that the strength of the gluonic

mixing term is comparable to the $\Delta I = 1$ mass term. If follows
that the eigenstates of the mass operator taking both the violation
of isospin and the gluonic mixing into account will not be close to      
being
eigenstates of the isospin symmetry.\\
\\
For the $\rho_{0}$--$\omega$ system
the mass operator takes the form:
\begin{equation}
\quad M = \left( \begin{array}{cc} M(v_u) & 0 \\ 0 & M(v_d)
\end{array} \right)\,
\end{equation}
\begin{displaymath}
\quad\quad\quad\quad\quad\quad \,+\,\,\tilde \lambda
\left( \begin{array}{cc} 1&1 \\1&1 \end{array} \right)\,.
\end{displaymath}
Using $M(v_{u})= M(\bar u u), M(v_{d})= M(\bar d d)$ and
$\tilde \lambda= 5.9$ MeV, we find
\begin{eqnarray}
|\rho_{0}\rangle = |\bar u u\rangle - |\bar d d\rangle  \nonumber
\end{eqnarray}
\begin{eqnarray}
&=& \mbox{cos}\, \alpha \,|\frac{1}{\sqrt 2}(\bar u u -\bar d d)\rangle
- \mbox{sin}\, \alpha \, |\frac{1}{\sqrt 2} (\bar u u +\bar d
d)\rangle\nonumber\\
&=& 0.997 |\frac{1}{\sqrt 2}(\bar u u -\bar d
d)\rangle + 0.071 |\frac{1}{\sqrt 2}(\bar u u +\bar d
d)\rangle\nonumber
\end{eqnarray}
\begin{eqnarray}
|\omega\rangle = |\bar u u\rangle + |\bar d
d\rangle \nonumber
\end{eqnarray}
\begin{eqnarray}
&=&\,\,\,\mbox{sin}\, \alpha \,|\frac{1}{\sqrt 2}(\bar u u -\bar d d)\rangle +
\mbox{cos}\, \alpha\, |\frac{1}{\sqrt 2}(\bar u u +\bar d
d)\rangle\nonumber\\
&=& -0.071 |\frac{1}{\sqrt 2}(\bar u u -\bar d
d)\rangle + 0.997 \,|\frac{1}{\sqrt 2}(\bar u u +\bar d
d)\rangle\nonumber
\end{eqnarray}                                                  
\begin{eqnarray}
M(\omega)-M(\rho_{0}) &=& \sqrt{(M(v_{u})-M(v_{d}))^{2}
+4\tilde \lambda^{2}} \nonumber \\ &=& 2.02\,\,\tilde \lambda
\end{eqnarray}
The mixing angle $\alpha$ discribing the strength of the
triplet--singlet mixing is about $-4.1^{o}$, i.e. a sizeable violation of
isospin symmetry is obtained. Neither is the $\rho_{0}$--meson an
isospin triplet, nor is the $\omega$--meson an isospin singlet.\\\\
The conclusions we have derived follow directly from the observed
smallness
of the gluonic mixing in the vector meson channel and the $m_{u}-m_{d}$
mass splitting, as observed e.g. in the mass spectrum of the
$K^{*}$--mesons.
Nevertheless they are in direct conflict with observed facts. According
to eq. (13), the probability of the $\rho_{0}$--meson to be an
$I=|\frac{1}{\sqrt 2}(\bar u u +\bar d d)\rangle$--state is
$\mbox{sin}^{2}\alpha \cong 0.51\%$. Taking into account the observed
branching ratio for the decay $\omega \rightarrow
\pi^{+}\pi^{-}$, BR $\cong(2.21\pm 0.30)\%$, this probability is bound
to
be less than $0.12\%$, in disagreement with the value derived above.
Obviously our theoretical estimate cannot be correct.\\
We consider the discrepancy described above as a serious challenge for
our understanding of the low energy sector of QCD. It arrives since the
strength of gluonic mixing is comparable to the estimated mass
difference
between the $|\bar u u\rangle$-- and $|\bar d d \rangle$--state. We can
envisage two possible solutions.\\
a)\, The strength of the gluonic mixing in the $1^{--}$--channel is much

larger than envisaged. This would lead to a substantial violation of the

Zweig rule and to a $\omega$--$\rho$ mass difference larger than
observed.
Thus an increase of $\tilde \lambda$ is excluded.\\
b)\, The mass difference $\Delta M = M(v_{d})-M(v_{u})$ must be smaller
than estimated above. In order to reproduce the observed branching ratio
for
the decay $\omega \rightarrow \pi^{+}\pi^{-}$, $\Delta M$ cannot exceed
$0.82$ MeV.\\
We believe that this is the correct solution of the problem, for the
following reasons. We consider the following two--point functions
\begin{eqnarray}
u_{\mu\nu}&=& \langle 0|\bar u (x)\gamma_{\mu}u(x)\,
\bar u (y)\gamma_{\nu} u(y)|0\rangle\,, \nonumber \\
d_{\mu\nu}&=& \langle 0|\bar d (x)\gamma_{\mu}d(x)\,
\bar d (y)\gamma_{\nu} d(y)|0\rangle \\
m_{\mu\nu}&=& \langle 0|\bar d (x)\gamma_{\mu}d(x)\,
\bar u (y)\gamma_{\nu} u(y)|0\rangle\,. \nonumber
\end{eqnarray}
The mixed spectral function $m_{\mu\nu}$ is expected to be
essentially zero in the low energy region, since the two different
currents can communicate only via intermediate gluonic mesons. In
perturbative QCD these states would be represented by three gluons. The
vanishing of $m_{\mu\nu}$ implies the validity of the Zweig rule. \\
The spectral functions $u_{\mu\nu}$ and $d_{\mu\nu}$ are strongly
dominated
at low energies by the $\rho_{0}$-- and $\omega$--resonances. The actual

intermediate states contributing to the two-point functions are $2
\pi$--
and $3 \pi$--states. However, a violation of the isospin symmetry due to

the $u-d$--quark mass splitting does not show up in the $\pi$-meson
spectrum. The $\pi^+- \pi^{\circ}$ mass splitting is due to the
electromagnetic interaction. It follows that resonant
$\left( 2 \pi \right)$ of $\left( 3 \pi \right)$ states, i. e. the
$\rho$--$\omega$--resonances, cannot display the effects of the isospin
violation either, and the mass difference
$\Delta M = M \left( v_d \right) - M \left( v_u \right)$
must be very small.\\
\\
Although the isospin symmetry is broken explicitly by
the $u-d$ mass terms, this symmetry violation does not show up in the
$\rho$--$\omega$ sector. The isospin symmetry breaking is shielded by
the
pion dynamics.\\
\\
Effectively the symmetry is restored by          
dynamical effects. Here the gluon anomaly plays an important role.
It might be that similar symmetry restoration effects are present in
other
situations, for example in the electroweak sector, which is sensitive to
the
dynamics in the TeV region.\\
\\

\end{document}